\documentclass[conference]{IEEEtran}
\usepackage{amsmath,amssymb,amsfonts, amsthm}
\usepackage{mathtools}
\usepackage{dsfont}
\usepackage{algorithmic}
\usepackage{algorithm2e}
\usepackage{float} 
\usepackage{afterpage}
\usepackage{graphicx}
\usepackage{textcomp}
\usepackage{xcolor}
\usepackage{comment}
\usepackage{gensymb}
\usepackage{pgfplots}
\usepackage{pgfplotstable}
\usepgfplotslibrary{fillbetween}
\usepackage[normalem]{ulem}
\usepackage{mdefs}
\usepackage{cite}
\usepackage{lipsum}
\usepackage{subcaption}
\usepackage[utf8]{inputenc}
\usepackage{hyperref}
\DeclareUnicodeCharacter{2061}{}
\usepackage[top=0.72in, bottom=1.0in, left=0.63in, right=0.66in]{geometry} 

\usepackage{algorithmic}
\usepackage{algorithm2e}

\floatstyle{ruled}
\newfloat{algorithm}{tbp}{loa}
\providecommand{\algorithmname}{Algorithm}
\floatname{algorithm}{\protect\algorithmname}

\def\BibTeX{{\rm B\kern-.05em{\sc i\kern-.025em b}\kern-.08em T\kern-.1667em\lower.7ex\hbox{E}\kern-.125emX}}

\def\argmin{\operatornamewithlimits{arg\,min}}

\def\GHz{~\mathrm{GHz}}
\def\MHz{~\mathrm{MHz}}
\def\Hz{~\mathrm{Hz}}

\def\m{\mathrm{m}}
\def\km{\mathrm{km}}

\def\s{~\mathrm{s}}
\def\dB{~\mathrm{dB}}
\def\dBm{~\mathrm{dBm}}

\def\U{\mathcal{U}}

\newtheorem{lem}{Lemma}

\title{Cooperative Sensing of Side Lobes Interference for mmWave Blockages Localization and Mapping}

\author{\IEEEauthorblockN{Hiba Dakdouk, Mohamed Sana, Benoit Denis}
\IEEEauthorblockA{CEA-Leti, Université Grenoble Alpes, F-38000 Grenoble, France\\
Email: \{hiba.dakdouk, mohamed.sana, benoit.denis\}@cea.fr}}

\newcommand{\titleheader}{This work has been accepted for publication in 2024 European Conference on Networks and Communications (EuCNC) \& 6G Summit}

\def\ps@IEEEtitlepagestyle{%
\def\@oddhead{\mbox{}\scriptsize \titleheader \rightmark \hfil }%
}
\makeatother

\begin{document}

\maketitle


\begin{abstract}
Radio localization and sensing are anticipated to play a crucial role in enhancing radio resource management in future networks. 
In this work, we focus on millimeter-wave communications, which are highly vulnerable to blockages, leading to severe attenuation and performance degradation. In a previous work, we proposed a novel mechanism that senses the radio environment to estimate the angular position of a moving blocker with respect to the sensing node.   
Building upon this foundation, this paper investigates the benefits of cooperation between different entities in the network by sharing sensed data to jointly locate the moving blocker while mapping the interference profile to probe the radio environment.
Numerical evaluations demonstrate that cooperative sensing can achieve a more precise location estimation of the blocker as it further allows accurate estimation of its distance rather than its relative angular position only, leading to effective assessment of the blocker direction, trajectory and possibly, its speed, and size.
\end{abstract}


\section{Introduction}

Millimeter-wave (mmWave) communication has emerged as a promising solution to meet the exponentially growing demand for high-speed data transmission and ultra-low latency applications. 
Operating at high frequencies (ranging, \egn, between $28$ and $300$ GHz), mmWave communications benefit from the availability of a large spectrum resource to enable next-generation wireless systems. 
However, mmWave technologies also come along with a critical challenge: communications at mmWave frequencies are highly vulnerable to blockages.
Blockages, caused by various environmental factors, such as buildings, trees, human body, or moving objects, can severely disrupt mmWave communication links, leading to link outages, reduced throughput, and degraded network performance. For example, penetration losses through the human body can vary between $20$ and $40\dB$, while attenuation through buildings can reach $40$ to $80\dB$.

Frequent interruptions and long-duration blockages can lead to severe degradation of end-user quality of service (QoS), requiring frequent handover procedures that affect network performance \cite{SanaHO2022}. 
Therefore, efficient blockage prediction mechanisms are essential to enable effective radio resource management (RRM). 
In this context, radio-localization and sensing techniques are envisioned as potential key enabler of intelligent RRM \cite{behravan2022positioning}.
Indeed, the sensing functionality can help measure or image the surrounding environment of communicating devices, enabling cognitive RRM and intelligent services \cite{liu2022integrated}. 
Here, we are interested in blockage prediction by sensing the surrounding environment of communicating devices to help anticipate on possible link failure and enable proactive RRM such as handover.
This fundamental research topic has attracted a particular attention from academia and industry \cite{oguma2016performance,charan2021vision,demirhan2022radar,marasinghe2021lidar,ali2019early,alrabeiah2020deep}.
One potential solution for predicting and preventing blockages in mmWave systems is to use in-band mmWave signal and data rate observations. 
By analyzing the fluctuations in the received signal strength (RSS) prior to the shadowing event, the authors in \cite{wu2022blockage} train a deep neural network to predict the future time at which a blockage will occur. 
However, the prediction accuracy decreases as the blocking object is further away from the mmWave link. 
Therefore, this approach is mostly effective when the blocking object is close to the mmWave communication beam. Another approach, proposed in \cite{koda2019handover}, uses deep reinforcement learning techniques to predict handover timings by analyzing fluctuations in the data rate prior to a shadowing event. 
This approach allows anticipating on the handover procedure before the degradation of user QoS. 
For the same purpose, the authors of \cite{hersyandika2022guard} propose an alternative approach that involves the use of an additional passive mmWave beam (called a ``guard beam") next to the main communication beam. 
This guard beam is intended to sense the environment by expanding the field of view of the base station (BS). 
By monitoring the fluctuations in the RSS from the guard beam, it is possible to detect a blocking object at an earlier stage.  
In a previous work \cite{sana2023sensing}, we proposed a mechanism that makes use of side-lobes information for passively and opportunistically sensing the surrounding of a dense mmWave network. 
Unlike the aforementioned studies, this approach leverages existing communication infrastructure, exploiting side lobes information to enable ubiquitous passive sensing. It exploits the spatial diversity of wireless network nodes and interference fluctuation in antenna side lobes caused by the existence of moving blockers in angular sectors around the communication link of concern. 

In this previous work \cite{sana2023sensing}, we consider applying this approach in an uplink-communication scenario where the base station is the sensing device and show that the proposed solution is capable of estimating the relative angle of the blocker \wrt the sensing node orientation and hence, \wrt to the communication link itself. The present work builds upon this approach while considering a downlink-communication scenario where the user equipment (UE) are the sensing devices. Most importantly, we propose a cooperative framework, sharing the information sensed by distributed UEs to jointly locate a moving blocker and map the interference profile, thus allowing a seamless sounding of the radio environment. 
This approach makes it possible to accurately predict the trajectory and obviously the velocity of the blocker, allowing early anticipation of blockage events and avoiding link outages by triggering, \egn, a handover procedure.

\section{System model}

\subsection{Network Model}
We consider a mmWave network composed of a set $\mathcal{B}=\{b_0,...,b_{M-1}\}$ of $M$ densely deployed BSs in a bi-dimensional Euclidean space of radius $R$ to provide service coverage to a set $\mathcal{U}=\{u_0,...,u_{K-1}\}$ of $K$ UEs. 
We assume BSs and UEs form two distinct homogeneous Poisson Point Processes (PPP) with densities $\lambda_b~[\m^{-2}]$ and $\lambda_u~[\m^{-2}]$ respectively such that in average, $\mathbb{E}[M] = \lambda_b \pi R^2$ and $\mathbb{E}[K] = \lambda_u \pi R^2$. 
We consider downlink communications. 
For simplicity, we assume each UE gets associated with the closest BS in an initial access phase where the UEs perform beam training and alignment mechanisms, configuring the appropriate beams, which exploit the maximum directivity gain \wrt serving BSs for the service phase.
In this dense network, a mobile and passive object, modelled as a cylindrical object of radius $r_B$ moves around with a relatively low velocity (\eg walking human, industrial or wheeled mobile robots), causing the blockage of interfering and direct communication paths.
In this work as in \cite{sana2023sensing}, we consider the existence of only one moving blocker, however our approach could be extended to detect multiple blockers. We leave this to future work.
Let $(x_B(t), y_B(t))$ denotes its instantaneous location \wrt UE $u_0$, referred to as the \emph{typical} UE and taken as the reference point in the following. In our previous work \cite{sana2023sensing}, we propose a novel approach for the passive sensing of such a moving object, by leveraging the interference perceived in the side lobes of the antenna radiation patterns.

\subsection{Sensing of side lobes interference}

\begin{figure}[!t]
    \centering
    \includegraphics{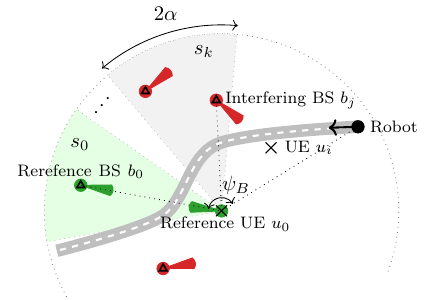}
    \caption{Network model viewed from UE $u_0$ perspective, consisting of 3 BSs interfering on the communication link ($b_0$ $ \rightarrow$ $u_0$). In this
    network, a mobile robot moves around causing blockages.}
    \label{fig:system_model}
\end{figure}

We assume spatial reuse of the spectrum across the network. For simplicity of analysis, we adopt a Friis path-loss model, where the received power $P_{\mathrm{Rx}}(t)$ is given as a function of the transmit power $P_{\mathrm{Tx}}$, and the distance $d$ between two nodes:
\begin{equation}\label{eq:channel-model}
    P_{\mathrm{Rx}}(t) = \chi(t) \zeta(t) P_{\mathrm{Tx}} G_{\mathrm{Tx}} G_{\rm H}(d) G_{\mathrm{Rx}}.
\end{equation}
Here, $\chi(t) = A \exp{(-0.125{r_B}^{-2}\psi_B(t)^2)}$ denotes the shadowing coefficient due to the passive object\footnote{We assume BSs and UEs are fixed and do not cause blockages.} moving around the corresponding link, where $A$ represents fully-shadowed attenuation and $\psi_B(t)$ is the relative blocker angle to the main communication link \cite{sana2023sensing}. 
Also, $\zeta(t)$ denotes the fading coefficient, $G_{\rm H}(d)$ denotes the path-loss gain, and $G_{\mathrm{Tx}}$ and $G_{\mathrm{Rx}}$ are the transmitter and receiver antenna gains respectively. 
In addition, let $I^{(i)}(\psi, t)$ denote the total interference perceived at time $t$ by UE $u_i$ as a function of interference signal angle of arrival (AoA) $\psi$ . When a blocker moves around the primary communication link of UE $u_i$, it shadows the perceived inference signals, causing the fluctuation of interference level\footnote{Obviously, the order of magnitude of such fluctuations is affected by several factors, including the blocker size, velocity, and trajectory.}, which we fully exploit to detect the moving blocker. To do so, the framework proposed in our previous work \cite{sana2023sensing} considers dividing the space around every UE $u_i$ into $n+1$ contiguous angular sectors, each of angular width $2\alpha=\dfrac{2\pi}{n+1}$, as represented in Fig. \ref{fig:system_model} that shows the sectorization \wrt UE $u_0$ as an example. This previous work introduces a novel metric called signal-to-sectored-interference-plus-noise ratio (S-SINR) $\gamma_{s_k}^{(i)}$, computed by aggregating the sum-interference level perceived in each sector $k$ as in \cite{sana2023sensing}: 
\vspace{-0.5em}
\begin{align}\label{eq:SINR}
    \gamma_{s_k}^{(i)}(t) = \frac{P_{\mathrm{Rx}}^{(i)}(t)}{I_{s_k}^{(i)}(t) + N_0 B},
\end{align}
where $P_{\mathrm{Rx}}^{(i)}(t)$ is the received power by UE $u_i$ from BS $b_i$ at time $t$, $B$ is the bandwidth, $N_0$ is the noise power spectral density, and the aggregated sector interference reads as
\begin{align}
    I_{s_k}^{(i)}(t) = \int_{s_k} I^{(i)}(\psi, t)\mathrm{d}\psi.
\end{align}

Consequently, for an observation window of size $\tau$, each UE $u_i$ computes a sensing matrix $\Lambda^{(i)}_{\tau,n}(t)$, an ordered-collection of S-SINR measurements in different sectors at different time-scale:
\begin{align}
\label{eq:sensingMatrix}
\boldsymbol{\Lambda}^{(i)}_{\tau,n}(t) = 
\begin{pmatrix}
\gamma^{(i)}_{s_0}(t) & \dots & \gamma^{(i)}_{s_n}(t)\\
\gamma^{(i)}_{s_0}(t-1) & \dots & \gamma^{(i)}_{s_n}(t-1)\\
\vdots  & \ddots & \vdots\\
\gamma^{(i)}_{s_0}(t-\tau) & \dots & \gamma^{(i)}_{s_n}(t-\tau)
\end{pmatrix},
\end{align}

The idea behind defining such matrix is that, the blockage of an interfering link around a certain UE in the sectorized regions induces fluctuations of the values of the sectors S-SINR values observed by UE $u_i$.
Applying a blind source separation technique based on singular value decomposition (SVD), the mechanism extracts the blocker signature matrix to estimate the active sector absolute angle\footnote{Specifically, $\theta_i$ corresponds to the estimated active sector absolute angle \wrt a global/shared angular reference system, given that the relative orientations of sensing devices are all known. We refer the reader to \cite{sana2023sensing}, for more details on the processing method $f(\cdot)$ applied for computing $\theta_i$.} $\theta_i(t) = f(\boldsymbol{\Lambda}^{(i)}_{\tau,n}(t))$ in which the blocker is thought to be located as shown in Fig. \ref{fig:sector_intersection}. This previous work \cite{sana2023sensing} focuses only on one sensing entity (which was a base station in the uplink scenario) and studies the blockage detection using one sensing matrix to estimate the blocker relative angle \wrt the sensing entity. In contrast, the present work considers a downlink scenario and cooperation between different UEs in the network by sharing their estimated angles, in order to enhance the accuracy and effectiveness of blockage localization.

\section{Proposed Cooperative Localization Scheme}\label{Sec:Proposed_Cooperative_Localization_Scheme}

\begin{figure}[!t]
    \centering
    \includegraphics[]{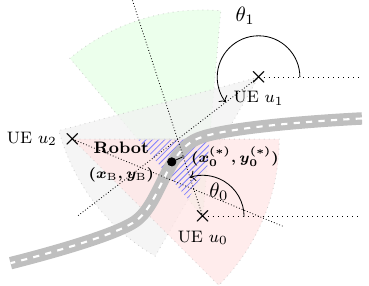}
    \caption{Cooperative localization scenario to localize robot blocker at $(x_B,y_B)$ with $N=3$ UEs. }
    \label{fig:sector_intersection}
\end{figure}

We now assume that each UE $u_i$ located at $(x_i,y_i)\in\mathcal{D}\subset \mathbb{R}^2$ has estimated the relative angle of the blocker located at $(x_B, y_B)$, as shown in Fig. \ref{fig:sector_intersection}, computing the active sector from its sensing matrix \eqref{eq:sensingMatrix} via the processing techniques proposed in \cite{sana2023sensing}. Let $\theta_i$ denote the estimated orientation of the active sector. Following \cite{sana2023sensing}, the accuracy of the angle estimation depends on the distance of the blocker to the sensing device, and is affected by multiple factors including false alarms or environment randomness such as channel fading. Therefore, we herein consider the cooperation between sensing devices via \eg information exchanges (i.e., with each other or with a central entity). Such data sharing enables to jointly locate the moving blocker and map the overall interference profile, hence contributing to accurately sound the radio environment. To do so, let us consider that the typical UE $u_0$ cooperates with $N$ neighboring UEs, which we denote with the set $\mathcal{N}_0$. By leveraging inter-user cooperation, as well as the natural spatial diversity allowed by sensing devices distribution, the typical UE can then compute the overlapping between different sectors and accordingly, pinpoint the absolute position of the blocker within the network, while mapping the interference conditions around each sensing node.

In the following, we describe our proposed approach to estimate the location of the blocker by computing the intersection of active sectors, as represented by the blue region in Fig. \ref{fig:sector_intersection}. This requires the knowledge of the detected active sector from each cooperating sensing device, as well as its absolute position. Thus, for a given UE $u_i$, 
we consider the quadruplet $(x_i, y_i, \theta_i, \alpha_i)$, where $2\alpha_i$ is the sector width, 
which depends \eg on the device capability and/or sensing procedure. However, for simplicity, in the following we consider $\alpha_i=\alpha,~\forall i$. 
From a single sensing node standpoint, when the blocker is detected in a sector, its estimated location $(\hat{x}_B, \hat{y}_B)$\footnote{To ease the reading and without loss of generality, we exclude the dependency on time.} is uncertain and can take any position along a line crossing $(x_i, y_i)$ with a slope $\tan(\theta_i + \epsilon)$, where $\epsilon$ is a random variable, which follows a uniform distribution $\epsilon \thicksim \U(-\alpha, +\alpha)$; \ien, 
\vspace{-0.5em}
\begin{align}
\exists \epsilon\in[-\alpha, \alpha], ~\text{s.t.}~ \hat{y}_B = y_i + \mathrm{tan}\left(\theta_i + \epsilon \right)(\hat{x}_B - x_i).
\vspace{-0.5em}
\end{align}
To compute the intersection of active sectors, we proposed to solve a least square estimation problem, following the approach proposed in \cite{rupp2019ls}. Specifically, for a given active sector with orientation $\theta_i$, assume $\epsilon_i\in [-\alpha, \alpha]$ is the localization error \wrt the relative sector orientation. Let $(x^{(\ast)}_i, y^{(\ast)}_i)$ define the closest point to $(\hat{x}_B, \hat{y}_B)$ along the line crossing $(x_i, y_i)$ with orientation $\theta_i + \epsilon_i$. We formulate the following optimization problem, which finds the location that minimizes the Euclidean distance between the estimated position of the blocker and the closest points $(x^{(\ast)}_i, y^{(\ast)}_i)$ of all cooperators:
\vspace{-0.5em}
\begin{equation}
\label{eq:estimated_coord}
    (\hat{x}, \hat{y}) = \argmin_{(x,y)} \sum_{u_i\in \mathcal{N}_0} \mathbb{E}_{\epsilon_i}\left[(x-x^{(\ast)}_i)^2 + (y- y^{(\ast)}_i)^2\right].
\end{equation}

The following Lemma shows how to compute the values of $\mathbb{E}_\epsilon[x^{(\ast)}_i]$ and $\mathbb{E}_\epsilon[y^{(\ast)}_i]$ (its proof is provided in the Appendix):

\begin{lem}
    \label{lem_locate}
    Given point $(\hat{x}, \hat{y})$, point $(x^{(\ast)}_i, y^{(\ast)}_i)$ along a line $(y-y_i)=tan(\theta_i+\epsilon)(x-x_i)$ closest to $(\hat{x}, \hat{y})$, where $\epsilon \thicksim \mathcal{U}(-\alpha, +\alpha)$, is such that:
    \begin{align}
        \mathbb{E}_\epsilon[x^{(\ast)}_i]=& \dfrac{1}{2\alpha} [A_{0,i} \hat{x} + A_{1,i} \hat{y} - A_{1,i}y_i + A_{2,i} x_i],\\
        \mathbb{E}_\epsilon[y^{(\ast)}_i]=&  \dfrac{1}{2\alpha} [A_{1,i} \hat{x} + A_{2,i} \hat{y} + A_{0,i}y_i - A_{1,i} x_i], 
    \end{align}
    where,
    \begin{align*}
        A_{0,i} &= \alpha + \dfrac{1}{2}\cos{(2\theta_i)}\sin{(2\alpha)},  \\
        A_{1,i} &= \dfrac{1}{2}\sin{(2\theta_i)}\sin{(2\alpha)},  \\
        A_{2,i} &= \alpha - \dfrac{1}{2}\cos{(2\theta_i)}\sin{(2\alpha)}.
    \end{align*}
\end{lem}

Then, with the data sensed by the set $\mathcal{N}_0$, minimizing the Euclidean distance in \eqref{eq:estimated_coord} simplifies to:
\begin{align}
    &(\hat{x}, \hat{y}) = \argmin_{(x,y)} \sum_{u_i\in \mathcal{N}_0}(x-\mathbb{E}_\epsilon[x^{(\ast)}_i])^2 + (y- \mathbb{E}_\epsilon[y^{(\ast)}_i])^2\\
    &\quad=  \argmin_{(x,y)} \dfrac{1}{2\alpha}\sum_{u_i\in \mathcal{N}_0} (A_{2,i}x -A_{1,i}y +A_{1,i}y_i -A_{2,i}x_i)^2 \nonumber\\
    & \quad \quad \quad \quad \quad \quad + (-A_{1,i}x + A_{0,i}y - A_{0,i}y_i+A_{1,i}x_i)^2, \nonumber 
\end{align}
whose solution, following the work in \cite{rupp2019ls}, is presented in the following lemma.

\begin{lem}
The minimum sum squared distance is found for point $(\hat{x}, \hat{y})$ that satisfies the following linear equation:
\begin{align}\label{eq:LSopt}
    &\begin{bmatrix}
        \sum_{u_i\in \mathcal{N}_0 } A_{2,i} & -\sum_{u_i\in \mathcal{N}_0 } A_{1,i} \\
        -\sum_{u_i\in \mathcal{N}_0 } A_{1,i} & \sum_{u_i\in \mathcal{N}_0 } A_{0,i}
    \end{bmatrix} 
    \begin{bmatrix}
        \hat{x} \\
        \hat{y}
    \end{bmatrix} \nonumber
    \\
   & = \begin{bmatrix}
        \sum_{u_i\in \mathcal{N}_0 }A_{2,i}x_i - A_{1,i}y_i \\
        \sum_{u_i\in \mathcal{N}_0 }-A_{1,i}x_i + A_{0,i}y_i
    \end{bmatrix}
\end{align}
\end{lem}

Hence, the position of the blocker is estimated by solving the linear equation \eqref{eq:LSopt} to obtain $(\hat{x}_B, \hat{y}_B)$. This can be efficiently computed at the typical UE or at a central orchestrator by gathering information $(x_i, y_i, \theta_i), ~u_i\in\mathcal{N}_0$ of neighboring sensing devices.


\section{Numerical Analysis}
We consider a network of $M$ BSs and $K$ UEs distributed in the space of a circular industrial environment of radius $R = 100~ \m$ according to homogeneous PPP with densities  $\lambda_b = 8\times10^{-4} ~\m^{-2}$ and $\lambda_u=2\times10^{-3} ~\m^{-2}$ respectively.
Each UE is associated to its closest BS and performs side lobes sensing.
A typical UE $u_0$, located at the center of the network, is taken as a reference. It cooperates with its $N$ neighboring UEs, via \eg a centralized entity or information exchange between each other to locate a moving blocker. Applying the proposed cooperative localization approach described in Sec.~\ref{Sec:Proposed_Cooperative_Localization_Scheme}, $u_0$ estimates the position of the blocker $(\hat{x}_B, \hat{y}_B)$ at every time $t$.
For all sensing UEs, we set the width of the angular sectors to $2\alpha=10\degree$ (\ie $n+1=36$ sectors overall, covering the 2D space), and the size of the observation window is set to $\tau=50\s$. 
We consider the same antenna and channel propagation model as in \cite{sana2023sensing}.
Other simulation parameters are presented in Table \ref{tab:sim_params}.

\begin{table}[!h]
\caption{Simulation parameters}
\centering
\begin{tabular}{c|c}
\hline
Parameters & Values \\
\hline
Carrier frequency & $28~ \GHz$\\ 
Bandwidth ${B}$ & $400~ \MHz$\\
Pathloss & $60.1 + 14 \log(d~[\km])$ \protect\cite{mudonhi2022mm}\\
Transmit power $P^{\mathrm{Tx}}$ (BS) & $33\dBm$\\
Noise power spectral density $N_0$ & $-174 \dBm/\Hz$ \\ 
Small-scale fading $\sim m$-Nakagami & $m=3$\\
{\rm Tx} beamwidth $\vartheta$ & 10°\\
{\rm Rx} beamwidth $\theta$ & 135°\\
$G_0(z)$& $\pi(21.32 z + \pi)^{-1}$ \protect\cite{Yang2018} \\ 
 $G_s^{\rm Tx}$& $G_0(\vartheta)$\\
 $G_m^{\rm Tx}$ &$G_0(\vartheta)\times10^{2.028}$  \\
$G_s^{\rm Rx}$& $0$\\
 $G_m^{\rm Rx}$ &$2G_0(\theta)\times10^{2.028}$  \\
Blockage Attenuation $A$ & $100\dB$\\
$\sigma_B$ & $\sqrt{8}~ r_B$
\end{tabular}
\label{tab:sim_params}
\end{table}

\begin{figure}[!t]
  \centering
  \begin{subfigure}[t]{0.4\textwidth}
    \includegraphics[width=0.85\textwidth]{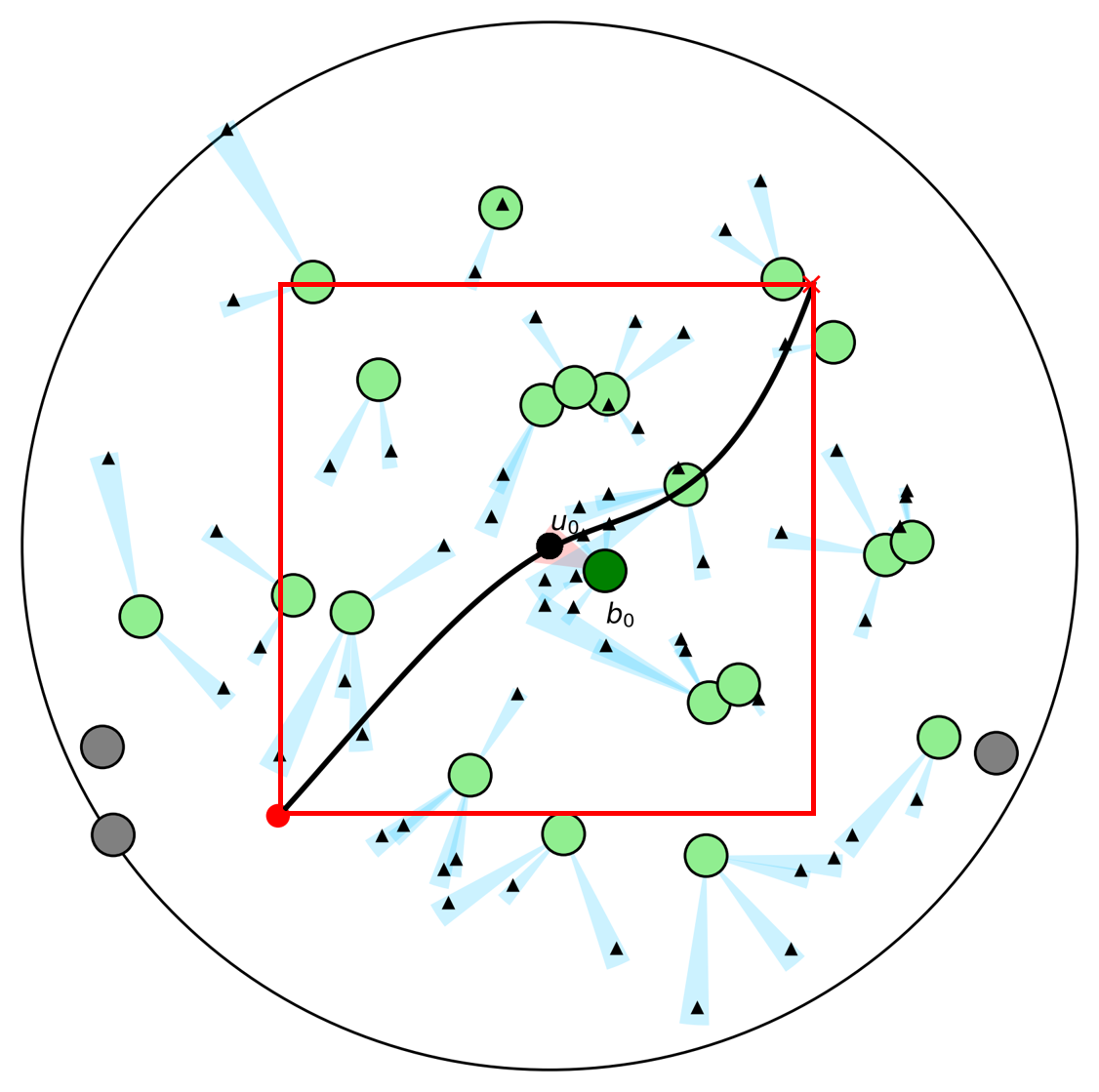}
    \caption{Deployment and blocker trajectory.}
    \label{fig:deployment}
  \end{subfigure}
  \begin{subfigure}[t]{0.4\textwidth}
  \centering
    \includegraphics[width=0.95\textwidth]{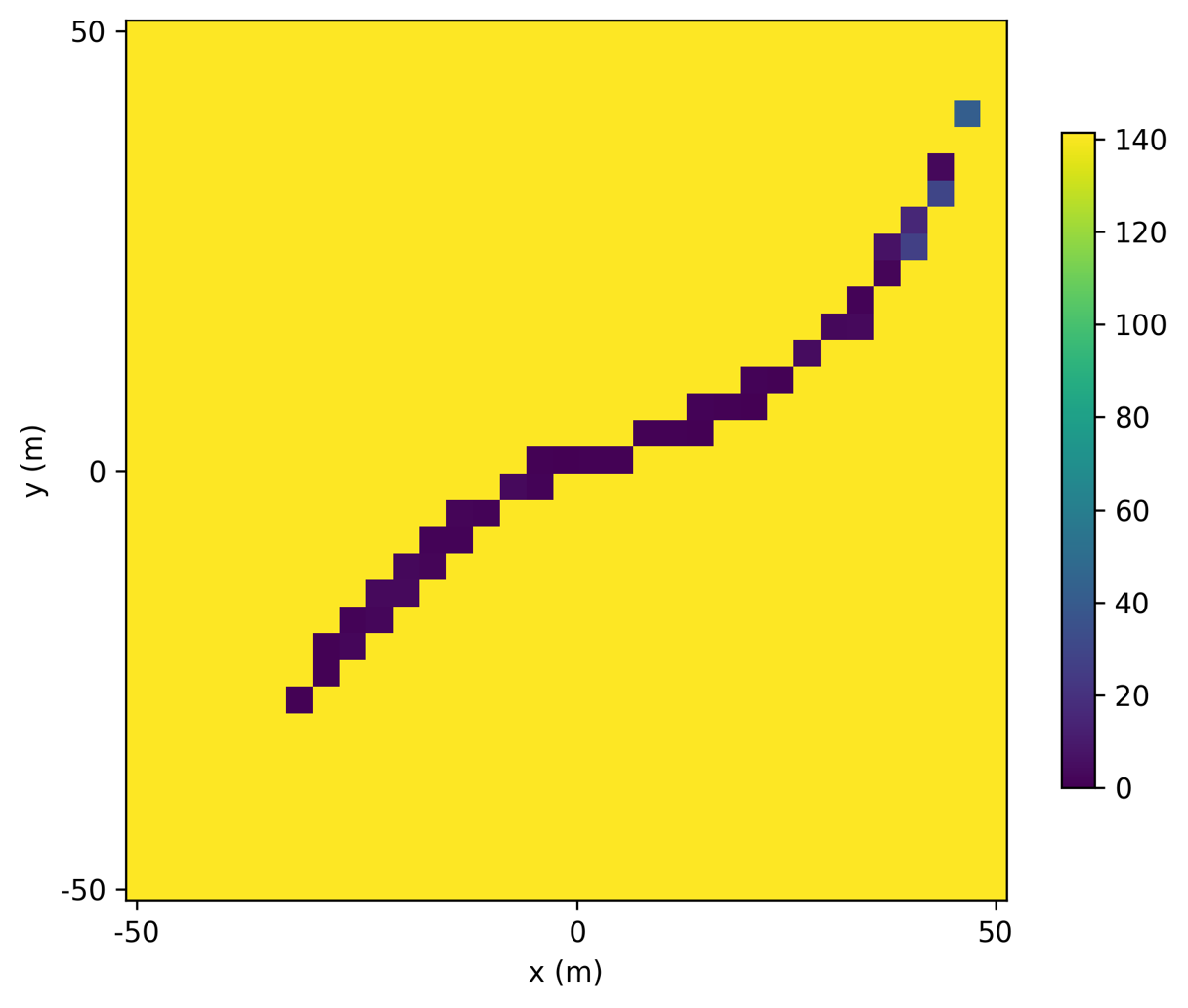}
    \vspace{-0.7em}
    \caption{Blocker Trajectory Estimation}
    \label{fig:trajectory}
  \end{subfigure}
  \caption{Example of blocker trajectory detection using our proposed cooperative side lobes sensing mechanism.}
  \label{fig:detection_example}
\end{figure}


\subsection{Detection of Blocker Trajectory}
For a random network deployment, we consider a mobile object of radius $r_B=1~\m$ moving along a random trajectory with an arbitrary velocity ($\sim [0.5, 2]~\m/s$). Fig. \ref{fig:deployment} shows an example of network deployment, where $M$ deployed BSs jointly provide service to $K$ UEs.
The reference UE $u_0$ cooperates with its $N=30$ closest UEs to estimate the position of the blocker at each time step.
To avoid cumbersome computations, we assume that the network area is partitioned into a mesh grid $\mathcal{G}$ consisting of $3~\m \times 3~\m$ cells. 
For a quantitative evaluation of the detection accuracy, we consider the following average location error, \ie the distance between the actual position of the blocker $(x_B, y_B)$ and the estimated position $(\hat{x}_B, \hat{y}_B)$, in each cell ($ c\in \mathcal{G}$):
\begin{align}
    {\delta} =  \mathbb{E}_{(x,y)\in c}\left[\sqrt{(x_B-\hat{x}_B)^2+(y_B-\hat{y}_B)^2})\right].
\end{align}
Fig. \ref{fig:trajectory} shows the mean estimation error in each cell the blocker passed by.
We can notice that the proposed method allows to localize the blocker and estimate its trajectory with very high accuracy, especially as it gets closer to the reference UE, as then it is more likely to cause blockages and get detected by the neighbor UEs of $u_0$. This information makes it possible to predict the trajectory and speed of the moving blocker, allowing for a plurality of resource management schemes, including anticipation of a blockage event and providing sufficient time to trigger proactive resource reallocation such as handover. 
\vspace{-0.5em}

\subsection{Localization accuracy vs blocker characteristics}
\begin{figure}[]
\centering
  \includegraphics[width=0.38\textwidth,height=9cm]{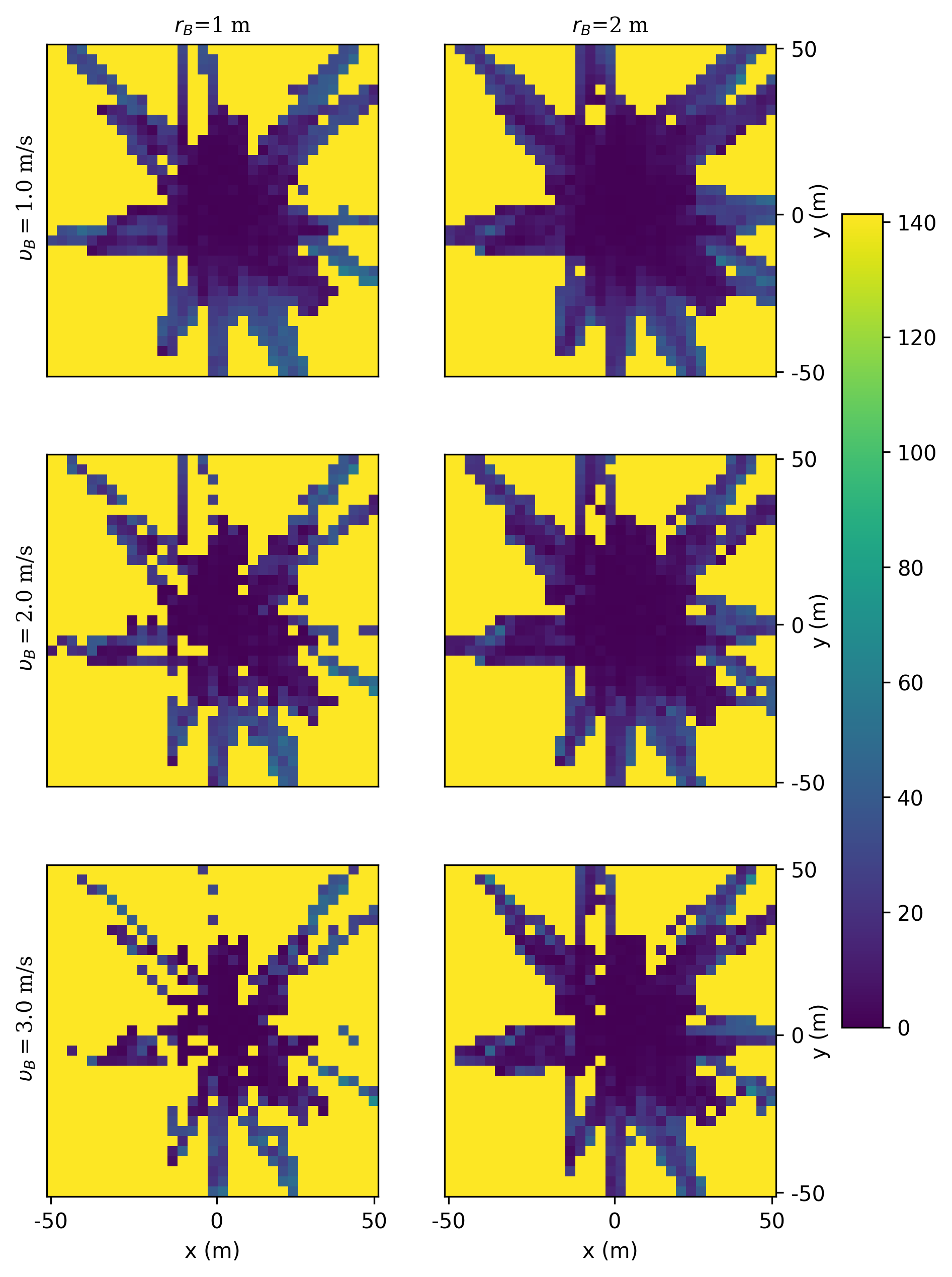}
  \caption{Localization error vs blocker speed and size. Here the number of cooperators is set to $N=10$.}
  \label{fig:error_vs_velocity}
\end{figure}

In this section, we assess the localization accuracy \wrt the blocker characteristics: size and velocity.
With the same deployment as in Fig. \ref{fig:deployment} and the same resolution grid $\mathcal{G}$, we consider a mobile object of radius $r_B$ scanning all the area in the red square zone of side $50~\m$ with a velocity $\upsilon_B$.
UE $u_0$ cooperates with its $N=10$ nearest UEs to localize the blocker. 
By varying the blocker velocity $\upsilon_B$ and radius $r_B$, we assess the localization error in each cell as shown in Fig. \ref{fig:error_vs_velocity}.
We can observe that as the blocker size increases, the accuracy improves as the blocker gets more detectable.
This is because a larger blocker is more likely to cause blockages of interference coming from different angles. 
We also observe that when the blocker is slower, it's more detectable by the UEs. 
This is due to the fact that with lower velocities, the blocker is more likely to be captured by the UEs at the time of sensing, making it possible to keep track of the blocker at every time step. 

\begin{figure}[t!]
\centering
  \includegraphics[width=0.38\textwidth,height=9cm]{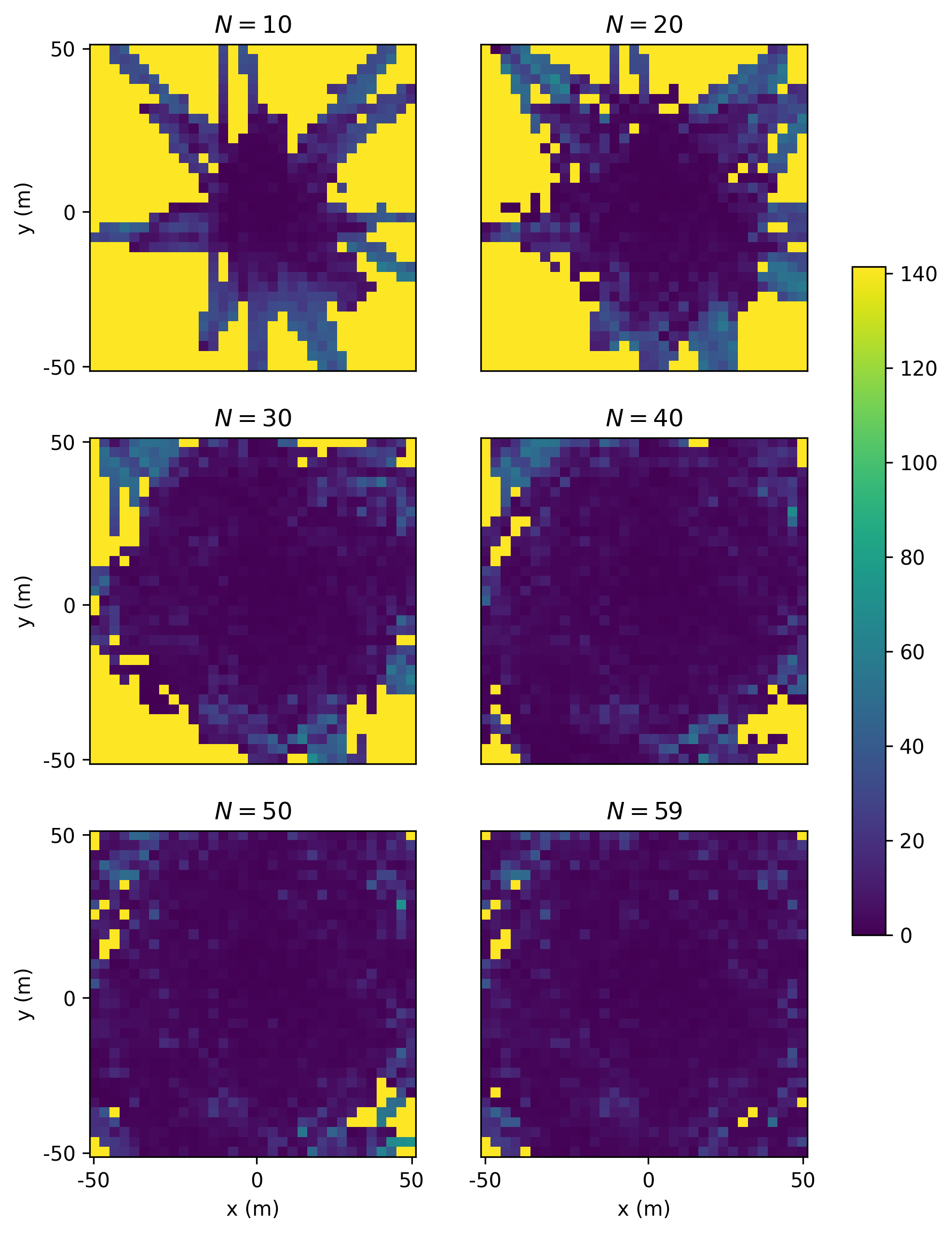}
  \caption{Localization error vs neighborhood size. Here the radius of the blocker is set to $r_B=1\m$.}
  \label{fig:error_vs_N}
\end{figure}
It is noticeable that the detection accuracy gets higher as the blocker approaches the center where $u_0$ is located. This is obviously because the cooperating UEs ($\mathcal{N}_0$) are chosen as the nearest to $u_0$, making the blocker more detectable in this region. To get a wider view of the network, it is however still possible to select farther UEs to share sensed information with.

\subsection{Localization error vs neighborhood size}

The size of the user neighborhood has a great impact on the detection range and the localization error of the proposed approach. 
Similar to the previous experiment, we present in Fig. \ref{fig:error_vs_N}, the impact of $\mathcal{N}_0$ on the estimation accuracy in each cell of the network.
The blocker moves with a velocity $\upsilon_B=1~\m/s$.
The reference UE $u_0$ cooperates with its nearest UEs in $\mathcal{N}_0$. The size $N$ of this neighborhood varies between $10$ and $59$. 
The results are presented in Fig. \ref{fig:error_vs_N}.
We can notice that as the number $N$ of cooperating nodes increases, the localization coverage is also improved since the area covered by the sensing UEs widens.
However, even with large neighborhood sets, the accuracy degrades as the blocker approaches the edge of the network, since it gets farther from the sensing UEs that cooperate in the location estimation step. 

As we are interested in locating mobile objects in millimeter-wave networks in order to anticipate and avoid blockages, the main region of interest, where it is crucial to detect blockages, is the region around the link to be protected against blockages (here around $u_0$).
We can then notice that even a small set of neighbors (\eg $N=10$) is sufficient to get a high estimation accuracy ($\delta\leq 1~\m$) in the main region of interest, which makes it possible to detect and avoid blockages with the least communication cost. 

\subsection{Discussion}
The preceding numerical experiments demonstrate  that our proposed cooperative approach can achieve a precise location estimation of a moving blocker.
Compared to previous studies, this approach leverages existing communication infrastructure without requiring any additional system, and it is capable to detect and track moving objects all around the sensing device, rather than being confined to a specific area. 
It is evident that our methodology holds significant promise for practical applications in anticipating and avoiding potential blockages in mmWave systems.
While this paper primarily focuses on detecting a single moving object, it could be further extended to detect multiple blockers by extracting the signatures of all moving objects responsible for SINR fluctuations and employing filtering techniques such as clustering to distinguish between them.


\section{Conclusion}
\label{sec:conclusion}

This paper proposes a novel collaborative framework that enables the sharing of sensed information between different UEs in the network.
With this new framework, we can locate and map accurately a moving blocker within the surrounding environment of a sensing UE. 
This is achieved by utilizing the fluctuations in the SINR values caused by the blocker's presence in the angular sectors around the communication link of interest. 
By using this collaborative approach, we can improve the accuracy and efficiency of the detection and localization of moving objects in mmWave networks, which is a crucial task for maintaining reliable mmWave communications.
Our numerical analysis shows that the proposed approach allows effective estimation of blocker position within the vicinity of the reference sensing device. 
It provides a level of localization accuracy, compliant with the anticipation of a blockage event, thus providing sufficient time to trigger proactive resource reallocation such as handover. 
 
Future work will consider data association and tracking problems in the same context in the presence \eg of multiple blockers. 
The results of this work shall also be used further to investigate smarter handover management mechanisms, and real-world experimentation is to be considered to validate the effectiveness and practical applicability of our methodology in dynamic and unpredictable environments. 
Furthermore, as our approach is flexible and scalable, it could be easily extended to more complex scenarios, \eg for 3D network operations \cite{maman2022coverage}.


\section*{Acknowledgment}
This work was supported by the french government under the Recovery Plan (CRIIOT project) and by the 6G-DISAC Project under the HORIZON program (no. 101139130).


\bibliographystyle{ieeetr}
\bibliography{biblio.bib}

\section*{Appendix}
\label{appendix}
\vspace{-0.3em}
Here, we provide the formal proof of Lemma \eqref{lem_locate}.
\vspace{-0.5em}
\begin{proof}
    Let $a_i = \tan(\theta_i + \epsilon_i)$, then, following \cite[Eq. 6-7]{rupp2019ls} we have :
    \begin{align}
        x^{(\ast)}_i &= \dfrac{1}{1+a_i^2} \hat{x} + \dfrac{a_i}{1+a_i^2}\hat{y} - \dfrac{a_i}{1+a_i^2}y_i +\dfrac{a_i^2}{1+a_i^2}x_i, \label{eq:xm*}\\
        y^{(\ast)}_i &= \dfrac{a_i}{1+a_i^2} \hat{x} + \dfrac{a_i^2}{1+a_i^2}\hat{y} + \dfrac{1}{1+a_i^2}y_i -\dfrac{a_i}{1+a_i^2}x_i. \label{eq:ym*}
    \end{align}

    Hence, we can compute the expected values of $x^{(\ast)}_i$ and  $y^{(\ast)}_i$ as follows:
    \begin{align*}
        &\mathbb{E}_\epsilon[x^{(\ast)}_i] =  \mathbb{E}_\epsilon \left [ \dfrac{1}{1+a_i^2} \hat{x} + \dfrac{a_i}{1+a_i^2}\hat{y} - \dfrac{a_i}{1+a_i^2}y_i +\dfrac{a_i^2}{1+a_i^2}x_i \right ] \\
        &\quad= \dfrac{1}{2\alpha}\int_{\epsilon=-\alpha}^{\epsilon=\alpha} \Bigg [ \dfrac{1}{1+\tan^2{(\theta_i + \epsilon)}} \hat{x} + \dfrac{\tan{(\theta_i + \epsilon)}}{1+\tan{^2(\theta_i + \epsilon)}}\hat{y} \\
        &\quad- \dfrac{\tan{(\theta_i + \epsilon)}}{1+\tan^2{(\theta_i + \epsilon)}}y_i -\dfrac{\tan^2{(\theta_i + \epsilon)}}{1+\tan^2{(\theta_i + \epsilon)}}x_i \Bigg] \,d\epsilon \\
        &\quad= \dfrac{1}{2\alpha} \left [A_{0,i} \hat{x} + A_{1,i} \hat{y} - A_{1,i}y_i + A_{2,i} x_i \right ] ,
    \end{align*}
    where,
    \begin{align*}
        A_{0,i} &= \int_{-\alpha}^{\alpha} \dfrac{1}{1+\tan^2{(\theta_i + \epsilon)}} \, d\epsilon= \alpha + \dfrac{1}{2}\cos{(2\theta_i)}\sin{(2\alpha)}\\
        A_{1,i} &= \int_{-\alpha}^{\alpha} \dfrac{\tan{(\theta_i + \epsilon)}}{1+\tan^2{(\theta_i + \epsilon)}}\, d\epsilon = \dfrac{1}{2}\sin{(2\theta_i)}\sin{(2\alpha)}\\
        A_{2,i} &= \int_{-\alpha}^{\alpha} \dfrac{\tan^2{(\theta_i + \epsilon)}}{1+\tan^2{(\theta_i + \epsilon)}} \, d\epsilon
        = \alpha - \dfrac{1}{2}\cos{(2\theta_i)}\sin{(2\alpha)}
    \end{align*}
    Similarly,
    \begin{align}
        \mathbb{E}_\epsilon[y^{(\ast)}_i] = \dfrac{1}{2\alpha} \left [ A_{1,i} \hat{x} + A_{2,i} \hat{y} + A_{0,i}y_i - A_{1,i} x_i \right ],
    \end{align}
\end{proof}
\end{document}